\documentclass[12pt]{iopart}

\usepackage{makecell}

\usepackage{iopams}
\usepackage[caption=false,labelformat=simple]{subfig}

\usepackage{mathrsfs}
\usepackage{graphicx}
\usepackage{color,ulem}
\usepackage{multirow}
\usepackage[table,xcdraw]{xcolor}
\usepackage[utf8]{inputenc}
\usepackage{lmodern}
\usepackage{textcomp}
\usepackage{amsthm}
\usepackage{bm}
\usepackage{float}
\usepackage{perpage}    

\usepackage[linesnumbered,ruled,vlined]{algorithm2e} 

\usepackage{footnote}
\MakePerPage{footnote}			
\makesavenoteenv{tabular}		

\usepackage{xcolor}		
\usepackage{hyperref}   

\usepackage{multirow}

\definecolor{orange}{HTML}{D1512D}

\newcommand{\R}{\mathbb{R}}
\newcommand{\N}{\mathbb{N}}

\newcommand{\argmin}{\mathop{\mathrm{arg\,min\,}}}

\begin{document}

\title[Iterative tomographic reconstruction in CBCT]{Iterative tomographic reconstruction with TV prior for low-dose CBCT dental imaging}

\author{Louise Friot-{}-Giroux, Françoise Peyrin, Voichita Maxim}

\address{Univ Lyon, INSA‐Lyon, Université Claude Bernard Lyon 1, CNRS, Inserm, CREATIS UMR 5220, U1294, F‐69100, LYON, France}
\ead{louise.friot-{}-giroux@creatis.insa-lyon.fr}
\vspace{10pt}
\begin{indented}
\item[]September 2022
\end{indented}

\begin{abstract}

\textbf{Objective} Cone-Beam Computed Tomography is becoming more and more popular in applications such as 3D dental imaging. Iterative methods compared to the standard Feldkamp algorithm have shown improvements in image quality of reconstruction of low-dose acquired data despite their long computing time. An interesting aspect of iterative methods is their ability to include prior information such as sparsity-constraint. While a large panel of optimization algorithms along with their adaptation to tomographic problems are available, they are mainly studied on 2D parallel or fan-beam data. The issues raised by 3D CBCT and moreover by  truncated projections are still poorly understood.

\textbf{Approach} We compare different carefully designed optimization schemes in the context of realistic 3D dental imaging. Besides some known algorithms, SIRT-TV and MLEM, we investigate the primal-dual hybrid gradient (PDHG) approach and a newly proposed MLEM-TV optimizer. The last one is alternating EM steps and TV-denoising, combination not yet investigated for CBCT. Experiments are performed on both simulated data from a 3D jaw phantom and data acquired with a dental clinical scanner.

\textbf{Main results} With some adaptations to the specificities of CBCT operators, PDHG and MLEM-TV algorithms provide the best reconstruction quality. These results were obtained by comparing the full-dose image with a low-dose image and an ultra low-dose image.

\textbf{Significance} The convergence speed of the original iterative methods is hampered by the conical geometry and significantly reduced compared to parallel geometries. We promote the pre-conditioned version of PDHG and we propose a pre-conditioned version of the MLEM-TV algorithm. To the best of our knowledge, this is the first time PDHG and convergent MLEM-TV algorithms are evaluated on experimental dental CBCT data, where constraints such as projection truncation and presence of metal have to be jointly overcome.
  
\end{abstract}

\vspace{2pc}

\submitto{\PMB}
%
%
%


\newpage



\section{Introduction}

Cone-Beam Computed Tomography (CBCT) is a 3D X-ray imaging technique where an object is reconstructed from a set of its 2D cone-beam projections acquired on 2D detectors. In dental imaging, CBCT enables to reduce the dose sent to the patient and to reduce the acquisition time compared to conventional CT scanners, making the use of such scanners much more frequent \cite{Liang2010}. However, as in other CT-based techniques, a trend in CBCT imaging is to reduce the ionizing radiation dose depositions, that come with subsequent undesirable effects for patients. 

Standard image reconstruction in CBCT is generally performed using analytical methods, providing the reconstructed image in a negligible amount of time. The FDK algorithm \cite{Feldkamp1984}, an heuristic extension of the 2D Filtered Back Projection (FBP) algorithm, is the most popular algorithm in CBCT despite the apparition of so-called cone-beam artifact when associated to a circular trajectory of the source. In addition, analytical methods relying on a continuous formulation of the direct problem, are quite sensitive to discretization conditions, and in particular to the number of projections. This therefore leads to reconstructed images likely to be affected by noise and various artifacts in low-dose conditions. Nevertheless, the FDK algorithm is still the most commonly employed method in dental CBCT, either used alone in the case of high-dose \cite{Baba2004} or associated to pre and/or post-processing to reduce some artifacts (metal in \cite{Ibraheem2012}, beam hardening and noise in \cite{Bayaraa2020}) and to improve the image quality.

In this context, iterative reconstruction methods are a flexible alternative to analytical methods. Basic iterative CT reconstruction methods include for instance, the conventional Algebraic Reconstruction Technique (ART) \cite{Gordon1970}, which is based on the Kaczmarz method and works ray by ray, or the Simultaneous Iterative Reconstruction Technique (SIRT) \cite{Gilbert1972} and Simultaneous Algebraic Reconstruction Technique (SART) \cite{Andersen1984} algorithms, that work on the whole volume and are more robust than ART. Statistical methods allow the inclusion of a priori information on the photon distribution statistics. Among them, the Maximum Likelihood Expectation Maximization (MLEM) algorithm \cite{Shepp1982} used in emission tomography has been adapted to transmission tomography in several works \cite{Lange1984,Vardi1985}. To speed up the reconstruction, the data can be divided into subsets leading to the so-called OSEM methods \cite{Hudson1994,Manglos1995}. 

However, in order to avoid the semi-convergent nature of most of these algorithms, the state-of-the-art solution is to include regularization in the form of prior information about the volume to be reconstructed, leading to the so-called model-based methods. Regularizers based on the $l_2$-norm have been widely used in the past. Sparsity-based regularizers such as total variation (TV), introduced for image denoising by Rudin, Osher and Fatemi \cite{Rudin1992} have demonstrated excellent results in many applications, including tomographic reconstruction \cite{Sidky2006,Banjak2018}. In \cite{Kolehmainen2003}, the authors applied statistical iterative methods to dental CT-scan, with a TV regularization. Some iterative algorithms already implemented on commercialized CT devices have shown very good results in terms of noise and artifacts suppression \cite{Widmann2017,Widmann2018}. However, they still remain barely used in practice for dental CBCT \cite{Kaasalainen2021}. 

Data-driven methods based on Deep Learning (DL) methods are also attractive for CT reconstruction problem. Various schemes have been proposed, most of the time based on post-processing and correction strategies to enhance a FDK reconstruction \cite{Jin2017,Park2020}. Another strategy consists in the direct resolution of the inverse problem as in \cite{Li2019}, where the network is conceived to transform sinogram data into images, or in \cite{Adler2018} where the physical model of the acquisition is included. Although DL methods can give impressive results on simulations, they usually need extensive training on representative datasets and require large datasets, especially in 3D CBCT. Moreover, some uncertainties subsist about their reliability for medical applications.

While a large panel of iterative algorithms are available, they have been mostly evaluated on 2D parallel or fan-beam data and rarely on 3D cone-beam projections. The specific geometry of CBCT coupled with dental imaging brings challenges in the reconstruction process that are not encountered in parallel beam models. On the one hand, despite hardware progresses made with the development of Graphic Processing Units (GPUs), computing resources are still an issue for iterative methods, especially in 3D CBCT. Finding a fast converging iterative scheme among the panel of available algorithms is thus relevant. On the other hand, there are also practical issues for applications to dental imaging like the problem of truncated projections since the detector size is smaller than the patient's head. Truncated projections lead to artifacts in the reconstructed image, creating bright bands that decrease the image quality and prevent a correct diagnostic. Several methods that mitigate truncation effects have been developed. Schemes extrapolating the projections, like including symmetric mirroring \cite{Ohnesorge2000} or water cylinder fitting \cite{Hsieh2004} have been proposed for analytical reconstructions. For iterative reconstructions, the most common strategy consists in increasing the reconstruction field of view \cite{Dang2016,Aootaphao2021}. Hence, the information issued from the truncated part is sent to the space provided by this extension of the matrix volume.

In this paper, we compare different iterative algorithms with TV regularization in order to reconstruct volumes from truncated and low-dose projections. 
We select some relevant algorithms from different families encountered in the literature and we identify their strengths and weaknesses, as well as ways to overcome the last ones. To the best of our knowledge, these methods have never been applied to dental CBCT data and some of them required adjustments. Although our comparison is not exhaustive, it should give a fair survey of the potential of such methods.
The SIRT-TV method from \cite{Banjak2018} consists in alternating SIRT and TV denoising steps. The algorithm is accelerated with a FISTA (Fast Iterative Shrinkage Thresholding Algorithm) step \cite{Beck2009}.
An important objective being dose reduction, along with SIRT which is representative of algebraic methods and is widely used in non-medical applications where computing time is less an issue, we also consider two algorithms designed for Poisson distributed data. In this case, the cost function is composed of a Kullback-Leibler data fidelity term and a TV semi-norm.
One of them is a Primal Dual Hybrid Gradient (PDHG) scheme \cite{Chambolle2011} particularized for tomographic reconstruction in \cite{Sidky2012}. As in \cite{Sidky2012}, we use the preconditioned version of PDHG from \cite{Pock2011} that provides faster convergence and is better adapted to complex projector operators such as the one from CBCT. Following \cite{Sidky2012}, we will call this algorithm KL-TV.
As in SIRT-TV, a step-alternating strategy can also be used for Poisson data \cite{Sawatzky2008,Yan2011}. These methods are a step forward compared to the Maximum A Posteriori (MAP) approach introduced in \cite{Green1990} which requires a very small regularization parameter, tends to blur the reconstruction and leads to numerical instabilities \cite{Panin1999,Persson2001}. We chose to use the step-alternating algorithm from \cite{Maxim2019b} which was proven to be convergent. The TV denoising is solved using the convex-duality principle of Fenchel-Rockafellar. To accelerate the denoising step and following \cite{Pock2011}, we also introduce a preconditioning that relies on the projection matrix. This algorithm will be referred as MLEM-TV.

In this work, since we focus on the suitability of these iterative reconstruction methods for low-dose dental CBCT imaging, we detail how to adapt them to the cone-beam geometry and to reconstruct from truncated projections. Our objectives are to: 1) evaluate the quality of the produced images; 2) evaluate the computing time and computing resources; 3) investigate potential of improvement. These carefully designed TV-based optimization schemes are then evaluated on an experimental CBCT acquisition.

This paper is organized as follows. In the second section, we introduce the notations and the basics of tomographic reconstruction. Then, we derive the framework of our method: the algorithms used are detailed in the second section, and the description of the data used for evaluation is given in the third section. In the fourth section, we present the results of the algorithms on phantom data to evaluate their robustness to noisy data, and, finally on experimental dental data. In the fifth section, we discuss the obtained results. Conclusions and perspectives are given in the last section.

\section{Methods}

\subsection{Tomographic reconstruction problem}

After traversing an attenuating medium, the intensity $\mathcal{I}$ of a X-ray beam having initial intensity $\mathcal{I}_0$ is modeled by the Beer-Lambert law. For a volume $f$ divided in $J=M\times N\times N$ voxels indexed by $j=1,\dots,J$, the Beer-Lambert law writes:
\begin{equation}
\label{eq:beer-lambert}
\mathcal{I} = \mathcal{I}_0 \exp(-\sum_j L_j f_j),
\end{equation}
where $f_j$ is the linear attenuation coefficient of the $j^{th}$ voxel and $L_j$ is the length of the intersection between the beam and the voxel. The volume $f$ can be determined when a sufficient number of beams indexed on $i=1,\dots,I$, are sent through the object.
We denote hereafter $a_{ij}$ the intersection length of the $i$th beam and the $j$th voxel, previously denoted $L_j$ in \eref{eq:beer-lambert}. The value:
\begin{equation}
\label{eq:log-BL}
p_i = -\log\left(\frac{\mathcal{I}}{\mathcal{I}_0}\right) = \sum_j a_{ij} f_j
\end{equation}
is the linear projection of the volume following the $i$th beam direction. The vector of projections will be noted hereafter $p$ and indexed on $i=1,\dots,I$. The tomographic problem is finally modelled by the following linear equation:
\begin{equation}\label{eq:Af=p}
Af = p,
\end{equation}
where $A=(a_{ij})$ is the system matrix and has dimensions $I\times J$. 

The measurement process is affected by several types of noises: electronic noise from the detector, which is supposed to be Gaussian, and Poisson photonic noise.
The lower is the initial X-ray source intensity, the higher is the noise in the acquired projections. In transmission tomography with high photon counts, the noise model is often assumed to be Gaussian for simplicity. For low-dose acquisitions the Poisson nature of the out-coming intensities has to be considered. The noise is thus not only additive and its modelling is a complex task \cite{Yu2012,Leuschner2021}.
In order to take this into account, our simulations include a Poisson-Gaussian mixture representative of both the low dose acquisition mode and of the electronic noise.

In this work, we consider reconstructions that are solutions of the following optimization problem:
\begin{equation*}
f^* = \argmin_{f} d(p,Af)+R(f)
\end{equation*}
where $d(p,Af)$ is the data consistency term and $R(f)$ a regularization term. The data consistency term constrains the reconstructed volume to fit the acquired data and allows, through the choice of the distance $d$, to add knowledge on statistical properties of the noise. The regularization term forces the solution to satisfy a priori information on the unknown object. A commonly used regularization term is the total variation (TV) semi-norm introduced in the next paragraph.

\subsection{Total Variation}~

Total variation regularization in tomographic reconstruction is a reference method when one has to deal with noisy projections and was previously used 
in particular for low-dose acquisition \cite{Persson2001, Sidky2006, Sawatzky2008, Anthoine2012, Yan2011}. For a function $f\in L^1(\Omega)$, with $\Omega$ an open subset of $\R^3$, the total variation semi-norm is given by :
\begin{equation}
\fl \quad \qquad TV(f) = \sup \left\{ -\int_\Omega f(x) \,\mathrm{div} \varphi(x) : \varphi \in C^1_C (\Omega,\R^3),\ \vert \varphi (x) \vert \leq 1\ \forall x\in\Omega \right\} \label{eq:TV_cont}
\end{equation}
where $C^1_C (\Omega,\R^3)$ is the space of compactly supported functions with continuous derivatives, and $\vert y\vert = \sqrt{y_1^2+y_2^2+y_3^2}$ for all $y=(y_1,y_2,y_3)\in\R^3$. 
This functional is finite if and only if the distributional derivative $Df$ of $f$ is a finite Radon measure on $\Omega$ (see for instance \cite{Chambolle2004}). Moreover, if $f\in W^{1,1}(\Omega)$, or equivalently $\nabla f\in L^1(\Omega)$, the total variation becomes $TV(f)=\int_\Omega \vert \nabla f(x) \vert dx$.

Hereafter we will use the discrete version of the total variation. Let $f$ be now a three-dimensional image, that is, an array of size $J=M\times N\times N$. We denote $X$ the Euclidean space $\R^{M\times N\times N}=\R^J$. The discrete gradient of $f$ is the array of elements $(\nabla f)_{i,j,k}= \left( (\nabla f)^1_{i,j,k},(\nabla f)^2_{i,j,k},(\nabla f)^3_{i,j,k} \right)$, where
{\setlength{\mathindent}{-0.4cm} \begin{equation*}
\qquad \begin{array}{lll}
(\nabla f)^1_{i,j,k} = \left\{ 
\begin{array}{ll} 
f_{i+1,j,k}-f_{i,j,k} & \mathrm{if}\  i<M \\ 
0 & \mathrm{if}\ i=M 
\end{array} \right.,
&
(\nabla f)^2_{i,j,k} = \left\{ 
\begin{array}{ll} 
f_{i,j+1,k}-f_{i,j,k} & \mathrm{if}\  j<N, \\ 
0 & \mathrm{if}\ j=N 
\end{array} \right.
\end{array} 
\end{equation*}}
\noindent and $\displaystyle (\nabla f)^3_{i,j,k} = \left\{ 
\begin{array}{ll} 
f_{i,j,k+1}-f_{i,j,k} & \mathrm{if}\  k<N, \\ 
0 & \mathrm{if}\ k=N 
\end{array} \right.$.

We denote $Y=X^3$ the space of gradients of three-dimensional images from $X$. The discrete total variation is then:
\begin{equation}
TV(f) = \sum_{i,j,k}\vert (\nabla f)_{i,j,k}\vert
\end{equation}
or, equivalently, 
\begin{equation}
\fl \quad \qquad TV(f)=\sup\left\{ \langle \varphi, \nabla f \rangle_Y : \varphi \in Y,\ \vert \varphi_{i,j,k}\vert \leq 1,\ i=1,\dots,M\ j,k=1,\dots,N \right\}
\end{equation}
where $\displaystyle \langle \varphi,\psi \rangle_Y = \sum_{i,j,k}\left( \varphi^1_{i,j,k}\psi^1_{i,j,k}+\varphi^2_{i,j,k} \psi^2_{i,j,k}+\varphi^3_{i,j,k} \psi^3_{i,j,k} \right) $. With the discrete divergence defined as :
{\setlength{\mathindent}{-0.8cm}\begin{equation*}
\qquad \begin{array}{lll}
(\mathrm{div}\varphi)_{i,j,k}= & 
\left\{ \begin{array}{cl}
\varphi^1_{i,j,k}-\varphi^1_{i-1,j,k}, &\mathrm{if}\ 1<i<M \\
\varphi^1_{i,j,k} & \mathrm{if}\ i=1 \\
-\varphi^1_{i-1,j,k} & \mathrm{if}\ i=M
\end{array} \right. +\ 
\left\{ \begin{array}{cl}
\varphi^2_{i,j,k}-\varphi^2_{i,j-1,k}, &\mathrm{if}\ 1<j<N \\
\varphi^2_{i,j,k} & \mathrm{if}\ j=1 \\
-\varphi^2_{i,j-1,k} & \mathrm{if}\ j=N
\end{array} \right. \\
& \qquad \qquad \qquad +\ \left\{ \begin{array}{cl}
\varphi^3_{i,j,k}-\varphi^3_{i,j,k-1}, &\mathrm{if}\ 1<k<N \\
\varphi^3_{i,j,k} & \mathrm{if}\ k=1 \\
-\varphi^3_{i,j,k-1} & \mathrm{if}\ k=N
\end{array} \right.
\end{array}
\end{equation*}}
\noindent we have $\langle\varphi,\nabla f\rangle_Y = -\langle \mathrm{div}\varphi ,f\rangle$ which leads to the discrete transcription of \eref{eq:TV_cont}.

\subsection{Reconstruction algorithms}

In this work, we then consider solving the TV regularization problem , expressed as
\begin{equation}
    f^* = \argmin_f d(p,Af) + \alpha TV(f)
    \label{eq:opt_pb}
\end{equation}
with $\alpha$ a positive parameter which controls the degree of smoothness. We consider that $d(p,Af)$ is either the least squares error norm $\Vert p-Af\Vert_2^2$ or the Kullback-Leibler distance
\begin{equation}
    KL(p,Af)=\sum_i p_i \ln p_i - p_i \ln(Af)_i + (Af)_i -p_i
    \label{eq:KL_def}
\end{equation}
The last one might be better suited for the high level of noise encountered in low-dose data.

Unlike parallel beam imaging, in CBCT the number of rays traversing a voxel is variable across the volume. More dose is given to the region of interest (ROI) since this is what we are interested in. 
This is reflected mathematically by the sensitivity, defined as:
\begin{equation}\label{eq:sensitivity}
    s= A^*\mathbf{1},
\end{equation}
where $\mathbf{1}$ is a column vector of ones and $A^*$, the adjoint (or transpose) of $A$, is the matrix of the back-projection operator. The term sensitivity is better known in emission tomography where it represents the probability for a photon to be detected somewhere.

We recall in the next subsections the non-regularized SIRT and MLEM algorithms and present the TV-minimization algorithms considered in this work.

\subsubsection{The SIRT algorithm}~

At each iteration, SIRT computes the weighted difference between the projections of the current volume and the acquired projections, then back-projects the result and subtracts it from the current volume:
\begin{equation}\label{eq:SIRT}
f^{(n+1)} = f^{(n)} + \lambda \frac{1}{A^*\mathbf{1}}A^*\left[ \frac{p-Af^{(n)}}{A\mathbf{1}} \right],
\end{equation}
with $\lambda>0$ the SIRT update step. SIRT can be seen as a weighted version of gradient descent for the minimization of the distance $\Vert Af-p\Vert^2_2$. The algorithm is slower but more stable than the Algebraic Reconstruction Technique (ART) \cite{Gordon1970} which uses a single projection at each iteration.

\subsubsection{The MLEM algorithm}~

The Maximum Likelihood Expectation Maximization algorithm (MLEM) and its variants are mainly used in emission tomography where the number of photons is low and they are detected individually. The EM (Expectation-Maximization) algorithm is an iterative technique used to maximize the log-likelihood function, which in tomographic reconstruction represents the probability that an image $f$ generates the measured projection data $p$. If we consider that the projections are Poisson distributed, the MLEM algorithm estimates the attenuation image by maximizing the log-likelihood 
\begin{equation}
l(f\vert p) = -\sum_{i=1}^I \sum_{j=1}^J a_{ij}f_j + \sum_{i=1}^I p_i \ln\left( \sum_{j=1}^J a_{ij}f_j \right) - \sum_{i=1}^I \ln(p_i!)
\label{eq:loglikelihood}
\end{equation}
under the positivity constrain on the coefficients of $f$. Maximizing \eref{eq:loglikelihood} is equivalent to minimizing
\begin{equation}
\label{eq:Lmlem}
 d(p,Af) = \sum_{i=1}^I \sum_{j=1}^J a_{ij}f_j - \sum_{i=1}^I p_i\ln\left( \sum_{j=1}^J a_{ij}f_j \right)
\end{equation}
which is the non-constant part of the Kullbak-Leibler distance \eref{eq:KL_def}. The EM algorithm applied to this minimization problem leads to the iterative scheme:
\begin{equation}
f_j^{(n+1)} = \frac{f_j^{(n)}}{s_j}\sum_{i=1}^I a_{ij}\frac{p_i}{\sum_{k=1}^J a_{ik}f_k^{(n)}}
\label{eq:MLEM}
\end{equation}
with $s_j = \sum_{i=1}^I a_{ij}$. Equation \eref{eq:MLEM} can be expressed with matrix multiplication and element-wise operations as:
\begin{equation}
f^{(n+1)} = \frac{f^{(n)}}{A^*\mathbf{1}} A^* \left[ \frac{p}{Af^{(n)}} \right]
\label{eq:MLEM_mat}
\end{equation}

MLEM algorithm is simple, efficient and allows to take into account the stochastic description of the problem. To further accelerate its convergence, Landweber-Kaczmarz acceleration techniques are usually applied and this leads to the ordered-subset expectation maximization or OSEM algorithm. As the number of iterations increases, high frequencies from the projections are progressively included and the image becomes less blurred and more precise. However, the algorithm has to be stopped before too many parasite high frequencies from noise are introduced. Finding a suitable number of iterations is usually a difficult task considering the balance between noise and precision. Stopping the iterations based on the discrepancy principle is a regularization method. In practice, the MLEM images are usually post-processed to remove the noise.

\subsubsection{The SIRT-TV algorithm}~

One approach to solve \eref{eq:opt_pb} with quadratic data fidelity is to perform TV denoising of the volume between successive SIRT steps. If $f^{(n+1/2)}$ is the volume after a SIRT iteration, the TV regularization aims at minimizing:
\begin{equation}
    \min_f \left\{ \frac{1}{2} \Vert f-f^{(n+1/2)}\Vert_2^2 + \alpha TV(f) \right\}.
\end{equation}
This minimization can be done by the Chambolle's algorithm \cite{Chambolle2004} as
\begin{equation}
f^{(n+1)} = f^{(n+1/2)} - \alpha \mathrm{div} \varphi^*\,,
\end{equation}
with $\varphi^*$ iteratively computed as the limit of : 
\begin{equation}
\varphi^{(k+1)} = \frac{\varphi^{(k)}+\tau\nabla(\mathrm{div}\varphi^{(k)}-f^{(n+1/2)}/\alpha)}{1+\tau \vert \nabla (\mathrm{div}\varphi^{(k)}-f^{(n+1/2)}/\alpha)\vert},
\end{equation}
starting from $\varphi^{(0)}=0$ and $\tau$ being an update parameter such that $0<\tau<1/12$ for three-dimensional images.

An acceleration method, the Fast Iterative Shrinkage-Thresholding Algorithm (FISTA) \cite{Beck2009a} was shown to speed up the convergence \cite{Banjak2016}. This technique consists in replacing the current volume $f^{(n+1)}$ by a linear combination
of $f^{(n+1)}$ and the previous value $f^{(n)}$ defined as:
\begin{equation}\label{eq:FISTA}
\bar{f}^{(n+1)} = f^{(n+1)} + \frac{t^{(n)}-1}{t^{(n+1)}}\left( f^{(n+1)}-f^{(n)} \right) .
\end{equation} 
The relaxation parameter $t^{(n)}$ is iteratively computed with $t^{(0)}=1$ and $\displaystyle t^{(n+1)}=\frac{1}{2}\left(1+\sqrt{1+4\left(t^{(n)}\right)^2} \right)$. 
The new volume $\bar{f}^{(n+1)}$ replaces $f^{(n)}$ in the next iteration, as input of the SIRT formula \eref{eq:SIRT}.

\subsubsection{The KL-TV algorithm}~

The algorithm proposed by Chambolle and Pock in \cite{Chambolle2011} aims at solving general optimization problems that can be written under the form 
\begin{equation}
\label{eq:primal}\min_f \left\{ F(Kf)+G(f) \right\}
\end{equation}
where $K$ is a linear operator, $F$ and $G$ are convex and possibly non-smooth. 
Sidky et al. adapted this algorithm to tomographic reconstruction in \cite{Sidky2012}. The optimization problem \eref{eq:opt_pb} with Kullback-Leibler data consistency term can be expressed in the form of \eref{eq:primal} with $F(Kf) = F_1(y)+F_2(z)$ and $G(f)=0$, where:
\begin{equation}
\begin{array}{l}K = \left( \begin{array}{l}
 A \\ \nabla 
\end{array}\right)\,, \\
y=Af,\ z=\nabla f\,, \\
F_1(y) = \sum_i \left[ y-p+p\ln p - p \ln (\mathrm{pos}(y))\right]_i + \delta_P(y)\,, \\
F_2(z)= \alpha \Vert ( \vert z \vert ) \Vert_1\,.
\end{array}
\end{equation}
In the equation above, $[pos(x)]_i=\max(0,x_i)$ and
\begin{equation}
    \delta_P(y) =\left\{ \begin{array}{cl}
    0 & \mathrm{if} \ y\ \mathrm{is \ positive}\\
    +\infty & \mathrm{otherwise}
    \end{array}\right.\,.
\end{equation}
Compared to the cost function in \eref{eq:opt_pb}, a positivity constraint on the projections has been added in order to enforce the positivity of the solution. In this paper, we use the pre-conditioned version from \cite{Pock2011,Sidky2012}, which allows significant acceleration of the convergence. The pseudo-code of the resulting KL-TV algorithm is given hereafter as algorithm \ref{algo:precondKLTV}. We denote by $\mathbf{1}$ a vector of ones with dimensions defined by the subscript. The subscript $I$, $P$ and $V$ denote respectively the image, the projections and the image gradient. $\vert M\vert$ is the matrix formed by taking the absolute value of each element of $M$.
Except for matrix-vector multiplications, all other operations are done element-wise. \vspace{0.4cm}

Compared to the original Chambolle-Pock algorithm, the preconditioned version has no parameters to tune once the relaxation parameter is fixed. All the other parameters have been replaced by the matrices $\Sigma_1$, $\Sigma_2$ and $T$.

\begin{algorithm}[H]
\DontPrintSemicolon
\KwIn{The acquired projections $p$, the TV parameter $\alpha$}
\KwOut{The reconstructed and denoised volume by KL-TV algorithm}
$\Sigma_1\gets \mathbf{1}_P/( A \mathbf{1}_I) ; \Sigma_2 \gets \mathbf{1}_V/(\vert\alpha\nabla\vert\mathbf{1}_I) ; T\gets \mathbf{1}_I/( A^*\mathbf{1}_P+\vert \alpha\mathrm{div}\vert \mathbf{1}_V) $\;
Initialize $u^{(0)}$, $z^{(0)}$ and $q^{(0)}$ to zero values \;
$\bar{u}^{(0)} \gets u^{(0)}$\;	
\While{$n \leq N$}{
	$y^{(n+1)} \gets \frac{1}{2} \left( \mathbf{1}_P + y^{(n)} + \Sigma_1 A\bar{f}^{(n)} - \sqrt{\left(y^{(n)} + \Sigma_1 A\bar{f}^{(n)} - \mathbf{1}_P\right)^2 + 4\Sigma_1 p}\, \right) $\;
	$z^{(n+1)} \gets \left( z^{(n)} + \alpha\Sigma_2 \nabla \bar{f}^{(n)} \right) / \max\left( \alpha \mathbf{1}_I,\vert z^{(n)} + \alpha\Sigma_2 \nabla \bar{f}^{(n)} \vert \right)$\;
	$f^{(n+1)} \gets f^{(n)} - T A^* y^{(n+1)} + T \mathrm{div}z^{(n+1)}$ \;
	$\bar{f}^{(n+1)} \gets 2 f^{(n+1)} - f^{(n)} $\;
	$\bar{f}^{(n+1)} \gets \mathrm{pos}\left(\bar{f}^{(n+1)}\right)$ \;
	$n\gets n+1$ \;
}
\Return{$\bar{f}^{(n+1)}$}\;
\caption{{\sc Pre-conditioned KL-TV}}
\label{algo:precondKLTV}
\end{algorithm}
\vspace{0.4cm}

\subsubsection{The MLEM-TV algorithm}~

The solution of the TV regularized reconstruction problem \eref{eq:opt_pb} with Kullback-Leibler data fidelity term can be obtained numerically with the EM algorithm \cite{Dempster1977}. At each iteration, two operations are completed: an (E) step which is the regular MLEM iteration,
\begin{equation}
f^{(n+1/2)} = \frac{f^{(n)}}{A^*\mathbf{1}} A^* \left[ \frac{p}{Af^{(n)}} \right]
\end{equation}
and a (M) step consisting into a TV denoising of the volume:
\begin{equation}
f^{(n+1)} \in \argmin_{f\in \R_+^J} \left\{ \langle f,s\rangle - \langle \ln(f),sf^{(n+1/2)}\rangle + \alpha TV(f) \right\} \label{eq:denoised_step}.
\end{equation}
The proof of the convergence and an efficient dual algorithm for the resolution of \eref{eq:denoised_step} are given in \cite{Maxim2019b}. By using the Fenchel-Rockafellar duality theorem, problem \eref{eq:denoised_step} can be reformulated through its dual and we get:
\begin{equation}
f^{(n+1)} = \frac{sf^{(n+1/2)}}{s+\alpha \mathrm{div}\varphi^*},
\end{equation}
with $\varphi^*$ iteratively computed starting from $\varphi^{(0)}=0$ and for $k\in\N$,
\begin{equation}
\varphi^{(k+1)} = \frac{\varphi^{(k)}-\tau z^{(k)}}{1+\tau\vert z^{(k)}\vert},\qquad \ z^{(k)}=\nabla\left( \frac{sf^{(n+1/2)}}{s+\alpha \mathrm{div}\varphi^{(k)}} \right). \label{eq:varphi} 
\end{equation}
Here $\tau>0$ is a minimization step. Let us note $s_{min}= \min\limits_j s_j$. To ensure the convergence of the algorithm, two constraints have to be verified:
\begin{itemize}
\item $\alpha < s_{min}/6$
\item $\tau < \alpha /L_h$ with $\displaystyle L_h = 12\alpha^2 \frac{\Vert sf^{(n+1/2)}\Vert_\infty}{(s_{min}-6\alpha)^2}$,
\end{itemize}
where $\Vert sf^{(n+1/2)} \Vert_\infty$ is the infinity norm of $sf^{(n+1/2)}$ seen as a one-dimensional vector.

One particularity of CBCT is that the sensitivity is far from being constant across the volume and one order of magnitude of difference can be observed between the values in the FOV and the values in the corners of the volume. The discrepancy becomes even more important when extensions of the volume are considered because of the truncated projections. As the minimization step $\tau$ depends on $s_{min}$, the convergence significantly slows down in the entire volume, even if the small values of the sensitivity are located in regions that are cropped after reconstruction. Following the idea of the preconditioned Chambolle-Pock algorithm, we test a new version of algorithm \eref{eq:varphi} where we replace $\tau$ with the matrix:
\begin{equation}
    T = 0.9\times \frac{(s-6\alpha)^2}{12\alpha sf^{(n+1/2)}}\,.
    \label{eq:T}
\end{equation}
This formulation allows faster convergence in the central part of the volume and thus for the region of interest.
As for SIRT-TV, a FISTA acceleration step \eref{eq:FISTA} is added to speed up the reconstruction.
A FISTA acceleration is also used during the computation of the sequence $\left(\varphi^{(k)}\right)_{k\in\N}$ from \eref{eq:varphi} to accelerate the convergence of the TV denoising. 

\section{Experimental data acquisition and phantom simulation}
\label{sec:data_description}

\subsection{3D CBCT acquisition geometry}

\begin{figure}[H]
    \centering
    \includegraphics[width=0.8\linewidth,trim=8cm 5cm 10cm 8cm,clip]{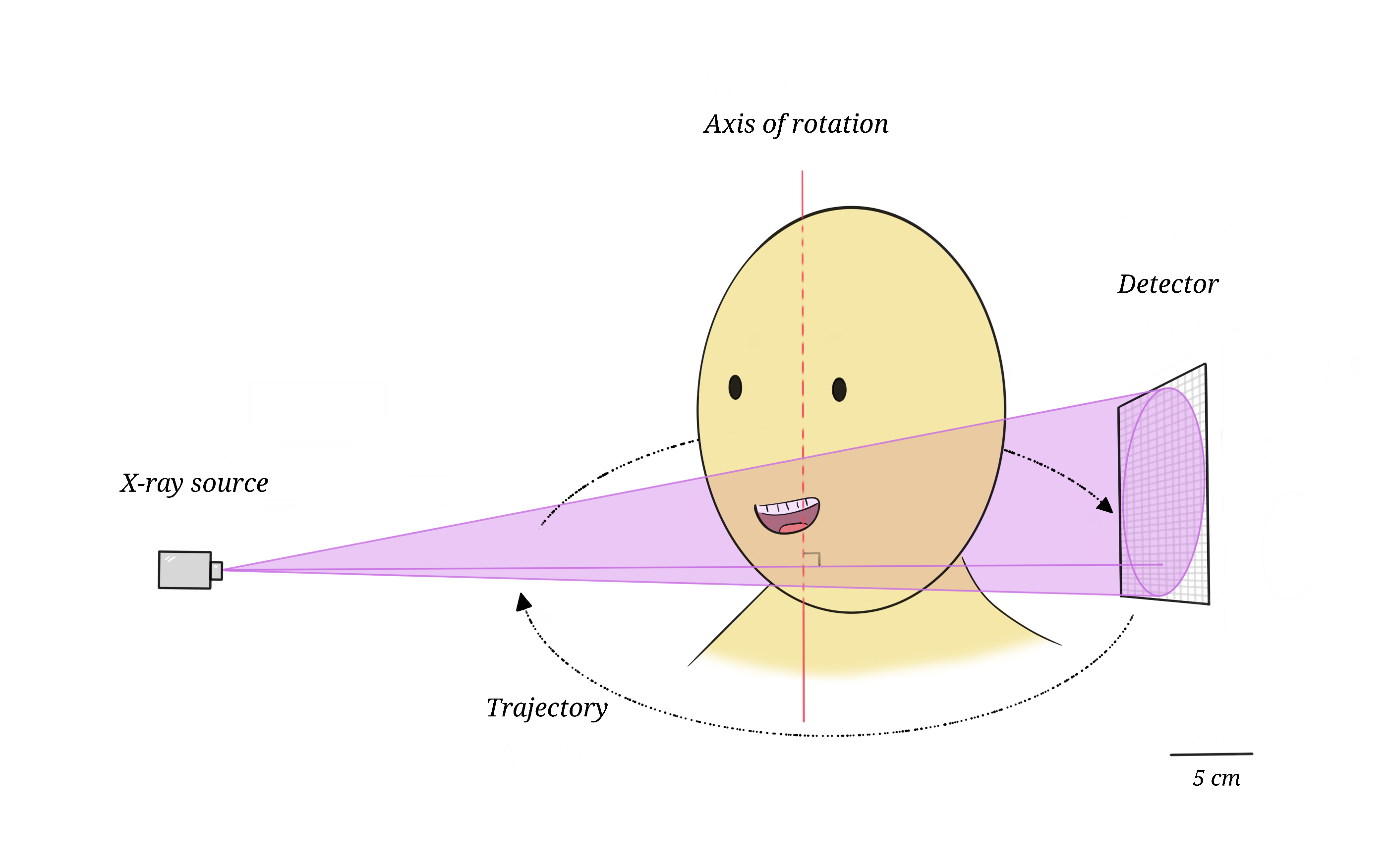}
    \caption{Diagram of the acquisition geometry.}
    \label{fig:geometry_scheme}
\end{figure}

The CBCT dental acquisition geometry is illustrated in \fref{fig:geometry_scheme}. The source and the detector rotate around the patient's head on a circular trajectory. The axis of rotation is located at 401.07 mm from the source and the distance source-detector is 564.30 mm. The size of the detector is $12\times 14$ cm, with $600\times 700$ pixels of size 200 µm. The projections are acquired with short scan conditions. 

\subsection{CBCT acquisition from a dental phantom}

A CBCT acquisition of an anthropomorphic head phantom delivered by W. Loy Gmbh\footnote{\url{https://www.loy-gmbh.de/produkt/dental-phantom/}} was realized with a Carestream Dental Scanner CS 8200 3D. A set of 155 projections was taken under a low-dose protocol, with 80 kV and 2 mA over 3.1s. 
To reduce the memory footprint during reconstruction, the projections were subsampled to $300\times 350$ pixels. The reconstructed volume is composed of $350\times 275\times 275$ cubic voxels with a size equal to 300 µm, giving a volume measuring $105\times82.5\times82.5$ mm$^3$.

Rather than adding a simulated noise on the real data, we have chosen to simulate an ultra-low dose acquisition by lowering the number of projections, which corresponds effectively to a dose reduction on scanners having pulsing capabilities. Both techniques are used in medical imaging to effectively reduce the dose received by the patient \cite{Liu2015,Humphries2019}. Thus, we consider an ultra low-dose version of these data, where 78 projections are regularly sampled from the low-dose data.

In order to mitigate the effect of truncation and following the idea from \cite{Maltz2007}, for the FDK reconstruction we extended the projections considering the FOV embedded into a cylinder. As in our experience this technique gave poor results with iterative reconstructions, we simply augmented the size of the reconstructed volume as in \cite{Dang2016}, by adding 150 voxels on each side. After reconstruction we truncated the volume to obtain only the ROI. 

\subsection{Three-dimensional jaw phantom}
The application of the algorithms to a numerical phantom allowed us to test their reliability and to adjust the different parameters.
We simulated a 3D jaw phantom inspired from the FORBILD phantom\footnote{\url{http://www.imp.uni-erlangen.de/phantoms/}}, containing in a cylinder representing the head of the patient some geometrical forms reproducing the spine, the jaw, as well as 31 teeth. Three teeth from the lower jaw have the density of the metal.
 The simulated volume has dimensions $105\times82.5\times82.5$ mm$^3$ sampled in $350\times275\times275$ voxels. A total of 78 projections with the same angles as for the experimental data (in ultra low-dose case) were calculated using the ASTRA Toolbox \cite{Van2016} in Python. In order to best represent the problem raised by the dental data at our disposal, the phantom projections were simulated with the same geometry as previously described, and the amount of noise to be added was determined from the experimental projections. 
To simulate the photonic noise, Poisson noise was added after scaling the projections by a multiplicative factor aiming to reach the same order of magnitude as in the experimental data. Finally, a Gaussian noise was added to simulate the electronic noise. 
The Gaussian noise level was estimated using the statistical relationship between the noise variance and the eigenvalues of the covariance matrix of patches extracted from the images \cite{Chen2015}. Negative values were replaced by zeroes.

\subsection{Algorithms implementation and evaluation metrics}

The algorithms have been implemented in Python and the ASTRA toolbox library has been used to compute the projection and backprojection operations. The reconstructions have been done on a Intel Xeon Gold 6226R processor and a Tesla V100 GPU.

As for the Shepp-Logan phantom the reference volume $f_{ref}$ is available, we evaluate the methods with the following metrics:
\begin{itemize}
	\item Normalized Root Mean Squared Error (NRMSE) is defined as follows: 
	$$ NRMSE(f,f_{ref}) = \frac{\Vert f - f_{ref}\Vert_2}{\Vert f_{ref} \Vert_2}$$
	The NRMSE computes the cumulative squared error between the reconstructed volume and the ground truth. A smaller NRMSE values points out that the reconstructed image is closer to the ground truth.
	\item The Peak Signal to Noise Ratio (PSNR) mirrors the difference in the noise level, in dB, between the two images.
	$$ PSNR(f,f_{ref}) = 10\log_{10}\left( \frac{\Delta^2}{\frac{1}{J}\sum_{j=1}^J(f_j-f_{ref,j})^2} \right), $$
	with $\Delta$ the data range of the true image, and $J$ the total number of voxels. A larger PSNR values indicates a better quality of the reconstruction.
	\item The Structural SIMilarity (SSIM) is used to measure the similarity between two images, using the luminance, the contrast and the structure of the two images:
	$$ SSIM(f,f_{ref}) = \frac{(2\mu_f\mu_{f_{ref}} +c_1)(2\sigma_{ff_{ref}}+c_2)}{(\mu_f^2+\mu_{f_{ref}}^2+c_1)(\sigma_f^2+\sigma_{f_{ref}}^2+c_2)}, $$
	with $c_1=0.01\times \Delta$ and $c_2 = 0.03\times\Delta$, $\Delta$ still being the data range of the true image.
\end{itemize}

We compare the convergence speed of the different algorithms in terms of both reconstruction iterations and time, by observing the decrease of the cost function. 

For the experimental data, we took as reference a normal dose FDK reconstruction of the phantom. However, the range of values of this reconstruction being different from ours, we could not use the same metrics as for the numerical phantom. We therefore used the Contrast-to-Noise Ratio (CNR) and the correlation with the reference image. For the calculation of the CNR, we compared the mean values on a patch inside a tooth ($\mu_{obj}$) and on a patch from the background ($\mu_{BG}$), with respect to the standard-deviation ($\sigma_{BG}$) of the noise in the background:
\begin{equation*}
    CNR(obj, BG) = 20\log_{10}\left( \frac{\vert \mu_{obj} - \mu_{BG}\vert}{\sigma_{BG}} \right).
\end{equation*}
To evaluate the correlation, we calculated the Pearson product-moment correlation coefficients between the images $f$ and $g$ having the same size as:
\begin{equation*}
    \mathrm{Corr}(f,f_{ref}) = \frac{\mathrm{cov}(f,f_{ref})}{\sigma_f \sigma_{f_{ref}}}, 
\end{equation*}
and with $\mu_f$ and $\mu_g$ being respectively the mean of $f$ and $g$,
\begin{equation*}
    \mathrm{cov}(f,f_{ref}) = \frac{1}{J-1} \sum_{j=1}^J \left[ (f_j-\mu_f)(f_{ref,j}-\mu_{f_{ref}}) \right].
\end{equation*}

\section{Results}

In this section, we analyze the performance of the different methods on the numerical jaw phantom in 3D and then we investigate issues specific to experimental data.

\subsection{3D jaw phantom }

The iterative methods require to set some parameters: the number of iterations ($N$), the number of iterations of the TV denoising step ($NTV$) and the regularization parameter $\alpha$. For SIRT-TV we also had to choose the value of $\lambda$ that we set to $0.8$. For larger values the algorithm was divergent.
All the parameters were chosen manually in the first place, by varying their values until we obtained a subjective good compromise between removing noise and conserving image features. In a second step, we tried to lower the number of iterations and the TV parameters, while keeping a reasonable image quality and a MSE value comparable to the one obtained with many iterations. Table \ref{tab:phantom_param} summarizes the final parameter values for the different algorithms.

\begin{table}[H]
\centering \begin{tabular*}{0.8\textwidth}{l @{\extracolsep{\fill}} cccc}
Method	 & \bf SIRT-TV	& \bf MLEM	& \bf MLEM-TV & \bf KL-TV \\ \hline
$N$		 & 400		& 200	& 400	& 500 \\ 
$NTV$	 & 20		&		& 20	&	  \\ 
$\alpha$ & $5\times 10^{-5}$ & & 0.1&0.1 \\ 
\end{tabular*}
\caption{Parameters used for the reconstruction of the 3D jaw phantom. $N$ is the number of reconstruction iterations, $NTV$ is the number of TV iteration if applicable and $\alpha$ is TV parameter used to tweak the importance of the regularization.}\label{tab:phantom_param} 
\end{table}

\begin{figure}[H]
\begin{minipage}{0.45\linewidth}
\centering \includegraphics[width=0.4\paperwidth]{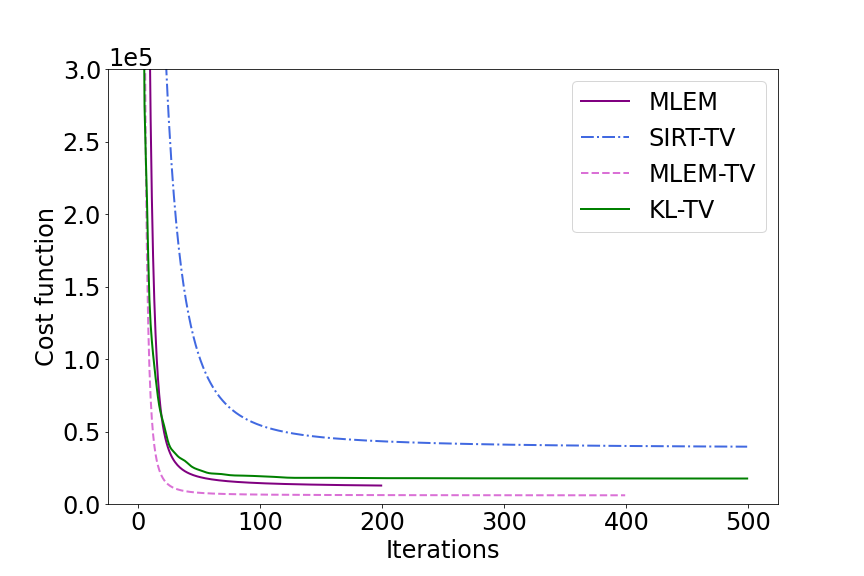}
\end{minipage}
\hspace{0.5cm}
\begin{minipage}{0.45\linewidth}
\centering \includegraphics[width=0.4\paperwidth]{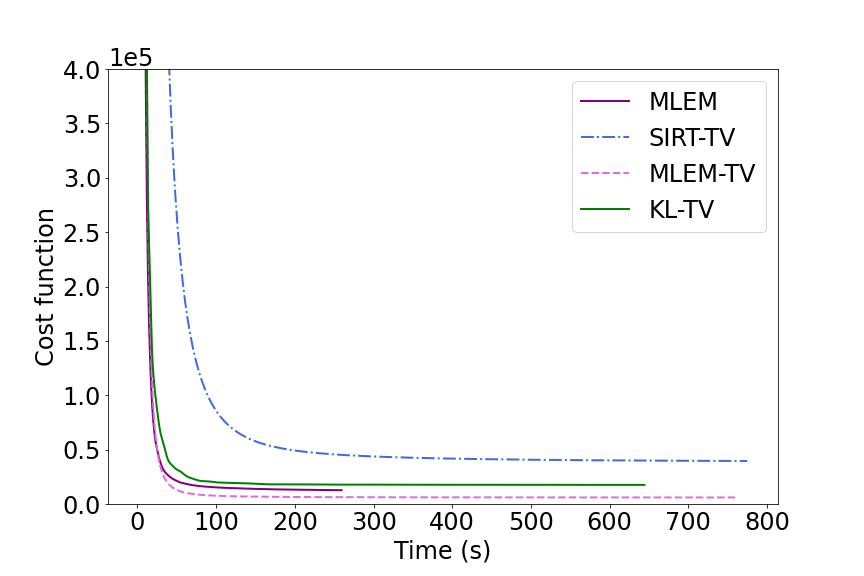}
\end{minipage}
\caption{Cost function evolution for the reconstruction of the phantom in low-dose mode. At the left, the cost function according to the iterations, and at the right, the cost function according to the time in seconds.}
\label{fig:phantom_cf}
\end{figure}

Figure \ref{fig:phantom_cf} shows the evolution of the cost functions as a function of the number of iterations and time for the different algorithms. We can notice the SIRT-TV is the slowest method to converge. 

The NRMSE, PSNR and SSIM values of the different reconstructions are shown in \tref{table:phantom_metrics}. All these metrics reflect the lower quality of the images provided by FDK for a small number of projections. 
MLEM is less affected by streak artifacts, which explains the much better values for the quality metrics compared to FDK reconstruction. The metrics for the three other methods are better. Regularized statistical methods obtain better results than SIRT-TV, suggesting that these algorithms are more suitable for the mixed Poisson-Gaussian noise model present in this study. They are roughly similar, with a slightly better performance for KL-TV.

\begin{table}[H]
\centering 
\begin{tabular*}{\textwidth}{l @{\extracolsep{\fill}} rrrrr}
\textbf{Metrics} & \textbf{FDK} & \textbf{MLEM} & \textbf{SIRT-TV} & \textbf{MLEM-TV} & \textbf{KL-TV} \\
\hline
\textbf{NRMSE} & 0.248 & 0.229 & 0.046 & 0.031 & \textbf{0.030} \\
\textbf{PSNR} & 39.684 & 41.354 & 50.504 & 54.328 & \textbf{57.216} \\
\textbf{SSIM} & 0.841 & 0.976 & 0.996 & 0.998 & \textbf{0.999} \\
\end{tabular*}
\caption{NRMSE, PSNR and SSIM for the reconstruction of the jaw phantom with the different algorithms.
}\label{table:phantom_metrics}
\end{table}

\begin{figure}[H]
\centering \includegraphics[height=8cm]{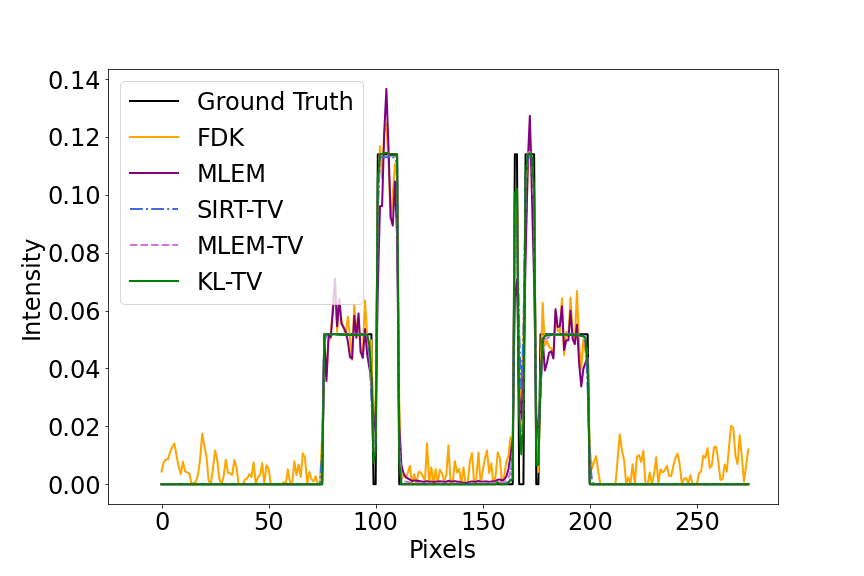}
\caption{Profile extracted from the reconstructed volumes.}
\label{fig:phantom_profile}
\end{figure}

A profile extracted from each volume and containing two teeth is shown in \fref{fig:phantom_profile}. We can note the noise in the FDK and MLEM methods and the positive impact of the regularization.

\begin{figure}[htp]
\centering
\subfloat[Upper jaw]{%
    \label{fig:phantom_ax175}
  \includegraphics[clip,width=\textwidth]{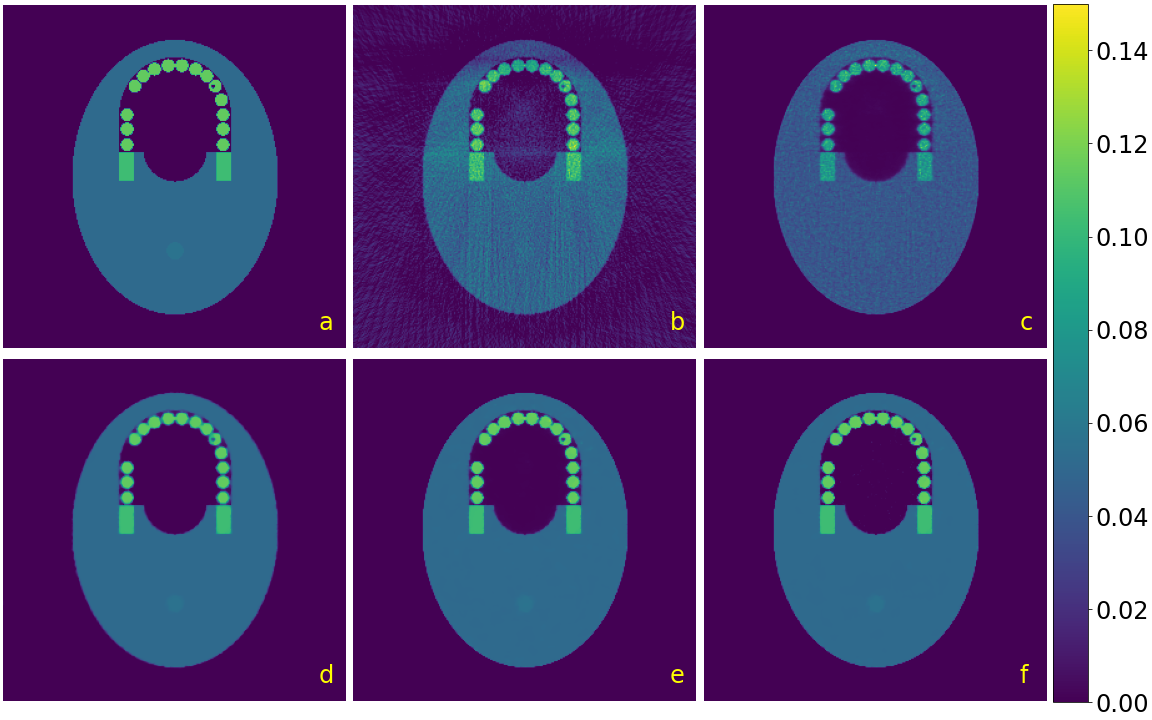}}

\subfloat[Lower jaw]{%
    \label{fig:phantom_ax2} 
  \includegraphics[clip,width=\textwidth]{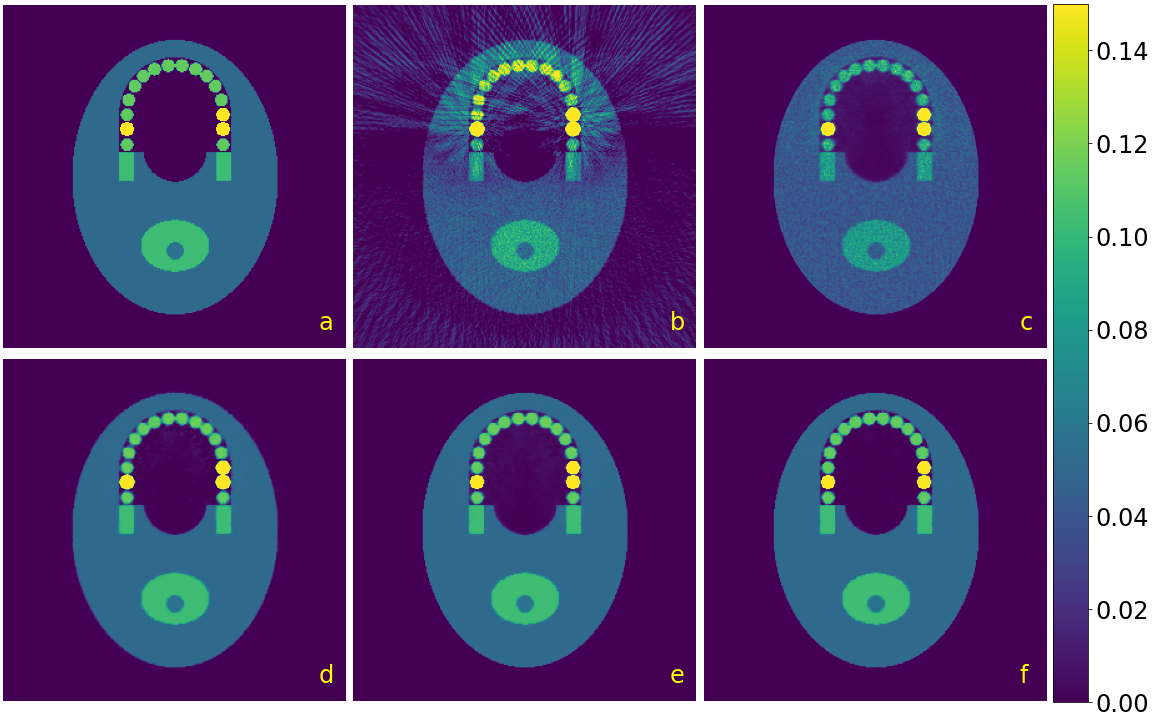}}
\caption{Axial slices from jaw phantom. (a) Ground truth, (b) FDK, (c) MLEM, (d) SIRT-TV, (e) MLEM-TV and (f) KL-TV.}
\label{fig:phantom_axial}
\end{figure}

\begin{figure}[htp]
\centering
\subfloat[Coronal slice from the phantom reconstruction.]{%
    \label{fig:phantom_cor}
  \includegraphics[clip,width=\textwidth]{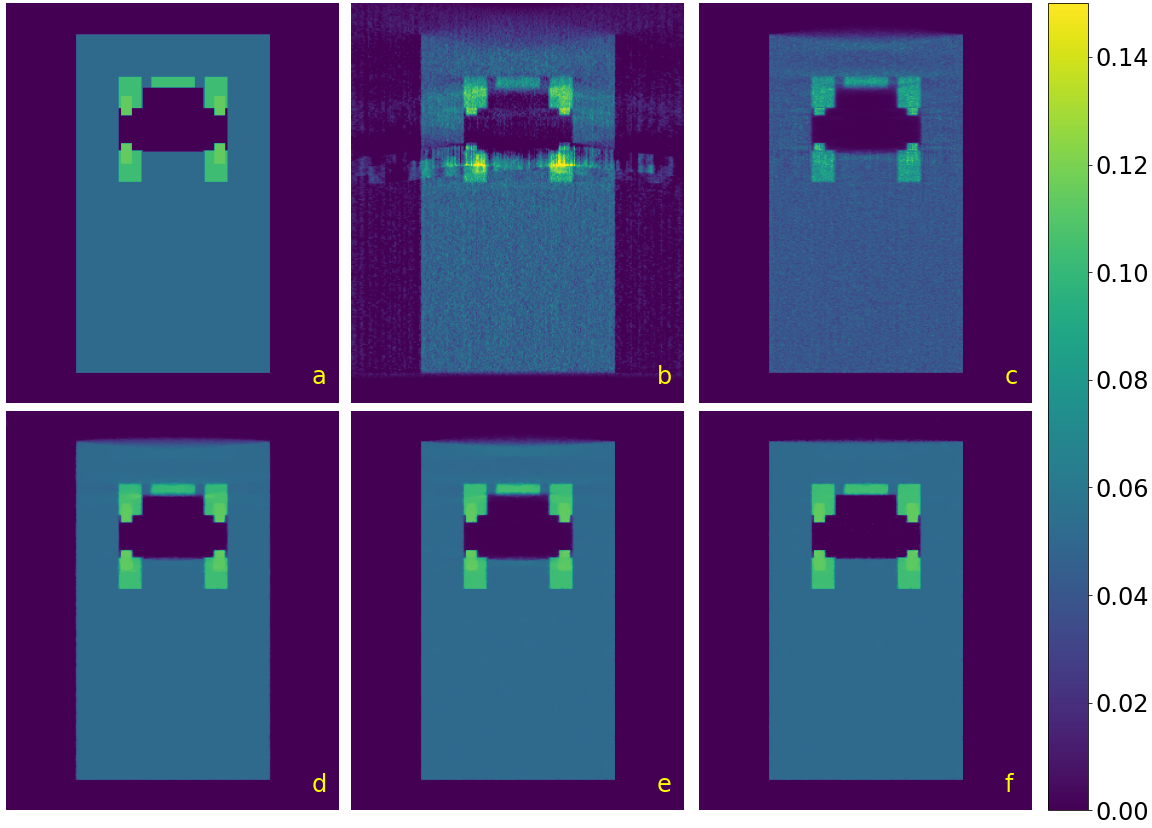}}

\subfloat[Zoom on the previous coronal slice.]{%
    \label{fig:phantom_cor_zoom} 
  \includegraphics[clip,width=\textwidth]{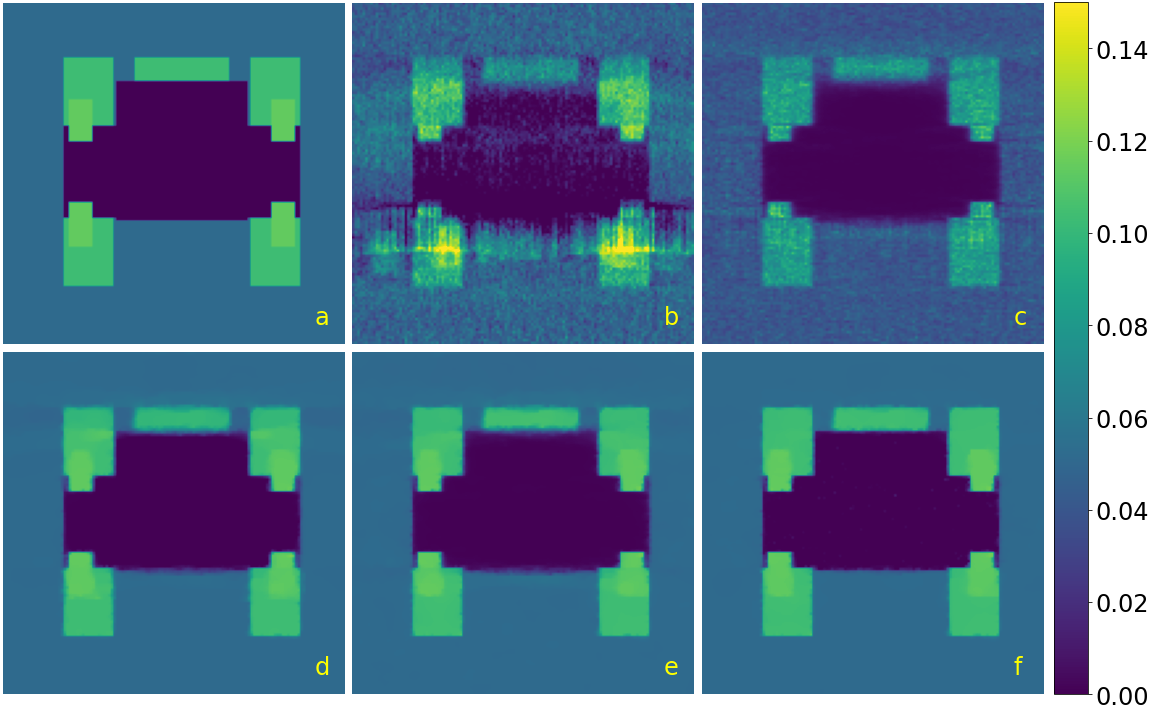}}
\caption{Coronal slice and zoom from jaw phantom. (a) Ground truth, (b) FDK, (c) MLEM, (d) SIRT-TV, (e) MLEM-TV and (f) KL-TV.}
\label{fig:phantom_coronal}
\end{figure}

Figure \ref{fig:phantom_axial} shows axial slices of the reconstructed volumes. From the axial slice shown in \fref{fig:phantom_ax175} it can be seen that all the methods succeeded to recover even the cavity inside the premolar from the jaw phantom, although the FDK and MLEM reconstructions are slightly noisy. Strike artifacts due to the limited number of projections can be observed in the FDK image. These artifacts are more visible in the second axial slice shown in \ref{fig:phantom_ax2}due to the presence of metal in this part of the jaw. 
We do not observe such artifacts in the iterative reconstructions.In particular, we can highlight the ability of MLEM to remove metal artifact, even without the use of regularization.
A coronal slice can be seen in figure \ref{fig:phantom_coronal}. The cone beam artifact visible at the top of the volume is particularly present in the reconstruction with FDK. The MLEM algorithm reduces it outside the object but the upper boundary of the phantom it is quite blurred.
Details from this slice are shown in \fref{fig:phantom_cor_zoom} where it can be noticed that the vertical lines are quite well reconstructed by the three iterative methods, while the horizontal lines are sharper with KL-TV. The zoom allows to better appreciate the effect of the TV regularization. All three methods achieve an accurate denoised reconstruction, even if SIRT-TV has less sharp edges than MLEM-TV and KL-TV.

In \tref{table:phantom_time}, the reconstruction times are shown. The MLEM algorithm is obviously faster than the other iterative algorithms since it does not do any regularization and the number of iterations is smaller. The regularized methods show comparable computation times with a slight disadvantage for MLEM-TV. The KL-TV method was faster to converge than MLEM-TV and SIRT-TV despite the larger number of iterations compared to the first one and equal for the second. This comes from the inner loop used to perform the denoising in the SIRT-TV and MLEM-TV algorithm.

\begin{table}[H]
\centering 
\begin{tabular*}{\textwidth}{l @{\extracolsep{\fill}} llll}
 \textbf{FDK} & \textbf{MLEM} & \textbf{SIRT-TV} & \textbf{MLEM-TV} & \textbf{KL-TV} \\
  [-0.45em]  & \footnotesize(200)  & \footnotesize(400-20)    &  \footnotesize  (400-20)  & \footnotesize (500) \\
\hline
00:00:02 & 00:04:19 & 00:12:55 & 00:12:40 & 00:10:47 \\
\end{tabular*}
\caption{Reconstruction time of the Shepp-Logan phantom for the different algorithms. The number of iterations used for the reconstruction and the TV denoising are reminded in the second line.}
\label{table:phantom_time}
\end{table}

This section studied the application of the chosen algorithms on the jaw phantom. This simple case allowed us to validate the algorithm on this geometry and helped us to choose the parameters for the next section.

\subsection{Experimental dental data}

We now apply the algorithms on experimental dental data: in a low-dose configuration which is currently used in dental imaging, and in an ultra low-dose configuration where only one out of two projections were used. 

Table \ref{tab:marcel_param} summarizes the parameters of the algorithms, in the low-dose and ultra low-dose case. We set the SIRT parameter to $\lambda=0.9$, compared to the phantom reconstruction, the details are more precise in the experimental data and we needed a higher parameter to speed up the convergence. As in the phantom case, we tried several reconstructions with different number of iterations and TV parameters, and kept the best in term of visualization and metrics. The same number of iterations was kept for the two cases. The TV parameter of SIRT was increased in the ultra low-dose reconstruction. The increase of this parameter in the MLEM-TV and KL-TV reconstructions resulted in too many details being lost. Similarly, increasing the number of iterations did not bring any improvement.

\begin{table}[H]
\centering \begin{tabular*}{\textwidth}{l @{\extracolsep{\fill}} lcccc}
&Method	 & \bf SIRT-TV	& \bf MLEM	& \bf MLEM-TV & \bf KL-TV \\ \hline
Low-dose &\bf $N$		 & 400		& 400	& 400	& 700 \\ 
&\bf $NTV$	 & 20		&		& 20	&	  \\ 
&\bf $\alpha$ & $1\times 10^{-6} $ & & 0.05&0.05 \\ 
Ultra low-dose &\bf $N$		 & 400		& 400	& 400	& 700 \\ 
&\bf $NTV$	 & 20		&		& 20	&	  \\ 
&\bf $\alpha$ & $2\times 10^{-6} $ & & 0.05&0.05 \\ 
\end{tabular*}
\caption{Parameters used for the reconstruction of the experimental data.}\label{tab:marcel_param} 
\end{table}

Compared to the phantom, more iterations had to be performed for most algorithms and especially for KL-TV. The number of iterations for the internal denoising loop was kept the same as its augmentation did not improve the results. Values of the TV parameter $\alpha$ are different from the jaw phantom reconstruction but remain in the same order of magnitude.

\begin{figure}[H]
\centering
\includegraphics[width=\textwidth]{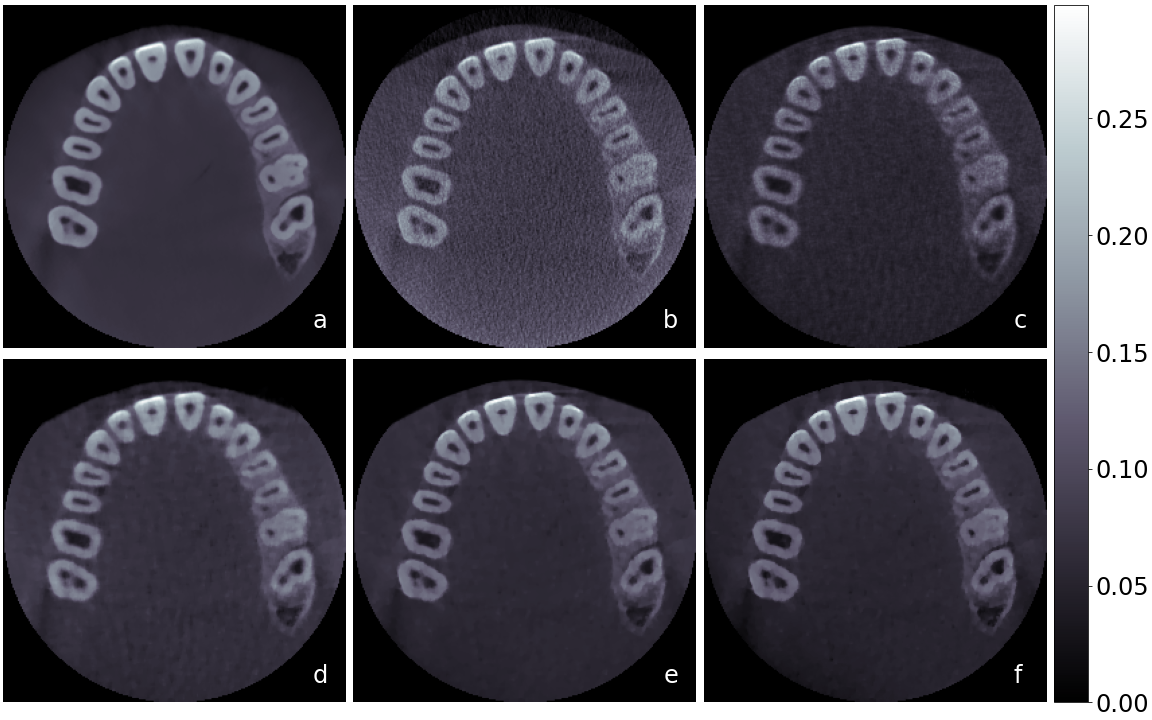}
\caption{Axial slice from the dental reconstruction, in low-dose. (a) Ground truth, (b) FDK, (c) MLEM, (d) SIRT-TV, (e) MLEM-TV, (f) KL-TV.}
\label{fig:mar_ax180}
\end{figure}

\begin{figure}[H]\centering
\includegraphics[width=0.7\textwidth]{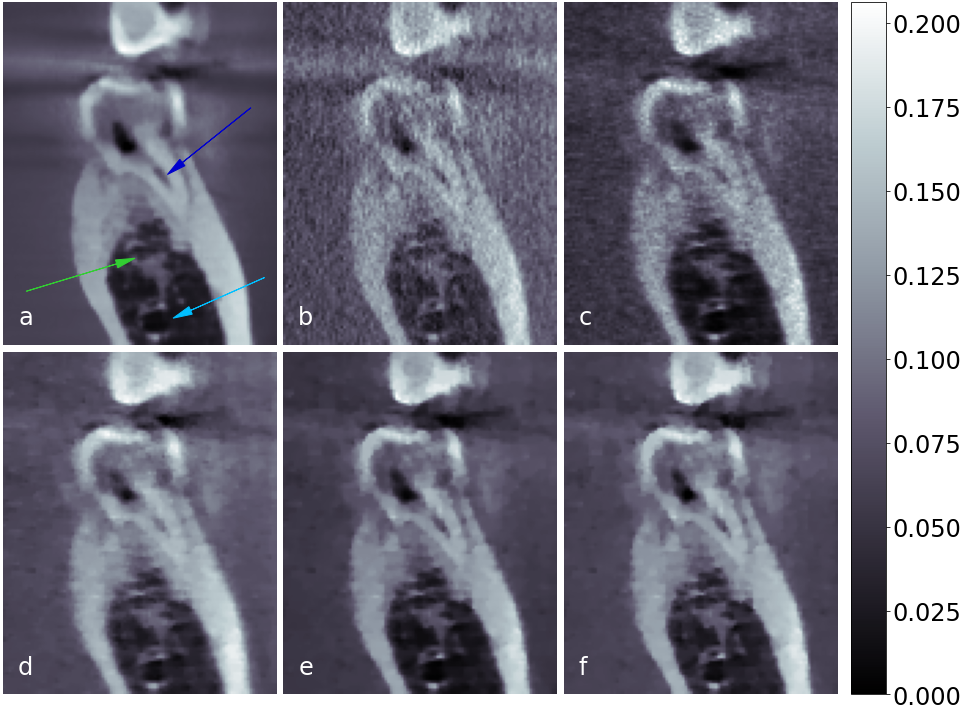}
\caption{Zoom on a tooth in a coronal slice, low dose acquisition. (a) Ground truth, (b) FDK, (c) MLEM, (d) SIRT-TV, (e) MLEM-TV, (f) KL-TV. The blue, green, light-blue arrows, respectively point to tooth canal, trabeculae bone and mandibular canal.}
\label{fig:mar_sag_zoom} 
\end{figure}

The values of the metrics used to evaluate the reconstructions are presented in \tref{tab:marcel_metrics}. Methods including TV regularization obtain the best results. The MLEM-TV reconstruction gets the best CNR, followed by the KL-TV one. We can notice that the CNR of ultra low-dose reconstructions is higher than that of low-dose. The CNR is a measure of contrast in an image. Since ultra low-dose reconstructions are noisier, the regularization parameter must be higher. 
However, as the TV denoised images tend to be piece-wise constant, an increase in the TV parameter also improves the contrast at the expense of reconstruction details.
The TV parameter was indeed increased for SIRT-TV. For MLEM-TV and KL-TV algorithms, this was not necessary: as the projections are two times less numerous, the value of $KL(Af,p)$ in the cost function is lower, so with the same TV parameter as in low-dose, more importance is given to the regularization.
The correlation has been calculated on three different regions: A is the entire volume, B is the 2D region shown in \fref{fig:mar_sag_zoom} and C is the background of the axial slice visible in \fref{fig:mar_ax180}. SIRT-TV performs better than the others on the entire volume (Corr A), with a tie with MLEM-TV in ultra low-dose. This result seems in contradiction with the visual observation of the reconstructions. This might be due to the large amount of pixels in the background compared to that in the teeth region, thus the background has a large influence in the correlation. 
When used to evaluate reconstruction of a tooth (Corr B), a better agreement with visual observation was obtained, as MLEM-TV and KL-TV reached a higher correlation with the ground truth; whereas when applied on a background area (Corr C), SIRT-TV has the greatest correlation, with a more significant gap compared to the total volume. To sum up, MLEM-TV and KL-TV gave the best results for the structures of interest, with a slight advantage to MLEM-TV.

\begin{table}[H]
\centering 
\begin{tabular*}{\textwidth}{l @{\extracolsep{\fill}} lccccc}
& \textbf{Metrics} & \textbf{FDK} & \textbf{MLEM} & \textbf{SIRT-TV} & \textbf{MLEM-TV} & \textbf{KL-TV} \\
\hline
Low-dose & \textbf{CNR} & 16.664 & 24.879 & 26.733 & \bf 28.163 & 27.290\\
& \textbf{Corr A} & $\ \,$0.933 & $\ \,$0.963 & $\ \,$\bf 0.974 & $\ \,$0.971 & $\ \,$0.960 \\
& \textbf{Corr B} & $\ \,$0.897 & $\ \,$0.937 & $\ \,$0.948 & $\ \,$\bf 0.954 & $\ \,$0.949 \\
& \textbf{Corr C} & $\ \,$0.950 & $\ \,$0.952 & \bf $\ \,$0.969 & $\,$ 0.962 & $\ \,$0.951 \\
Ultra low-dose$\ \ $ & \textbf{CNR} & 13.999 & 25.798 & 27.275 & \bf 30.182 & 29.308 \\
& \textbf{Corr A} & $\ \,$0.921 & $\ \,$0.960 & $\ \,$\bf 0.968 & $\ \,$\bf 0.968 & $\ \,$0.953 \\
& \textbf{Corr B} & $\ \,$0.817 & $\ \,$0.932 & $\ \,$0.929 & $\ \,$\bf 0.952 & $\ \,$0.950 \\
& \textbf{Corr C} & $\ \,$0.921 & $\ \,$0.946 & \bf $\ \,$0.968 & $\,$ 0.957 & $\ \,$0.943 \\
\end{tabular*}
\caption{CNR and correlation of experimental data, in low-dose and ultra low-dose cases. Corr A is the value of the correlation on the entire volume, while Corr B is calculated on a 2D region containing a tooth and Corr C on a background region.}
\label{tab:marcel_metrics}
\end{table}

\begin{figure}[H] \centering
\includegraphics[width=\textwidth]{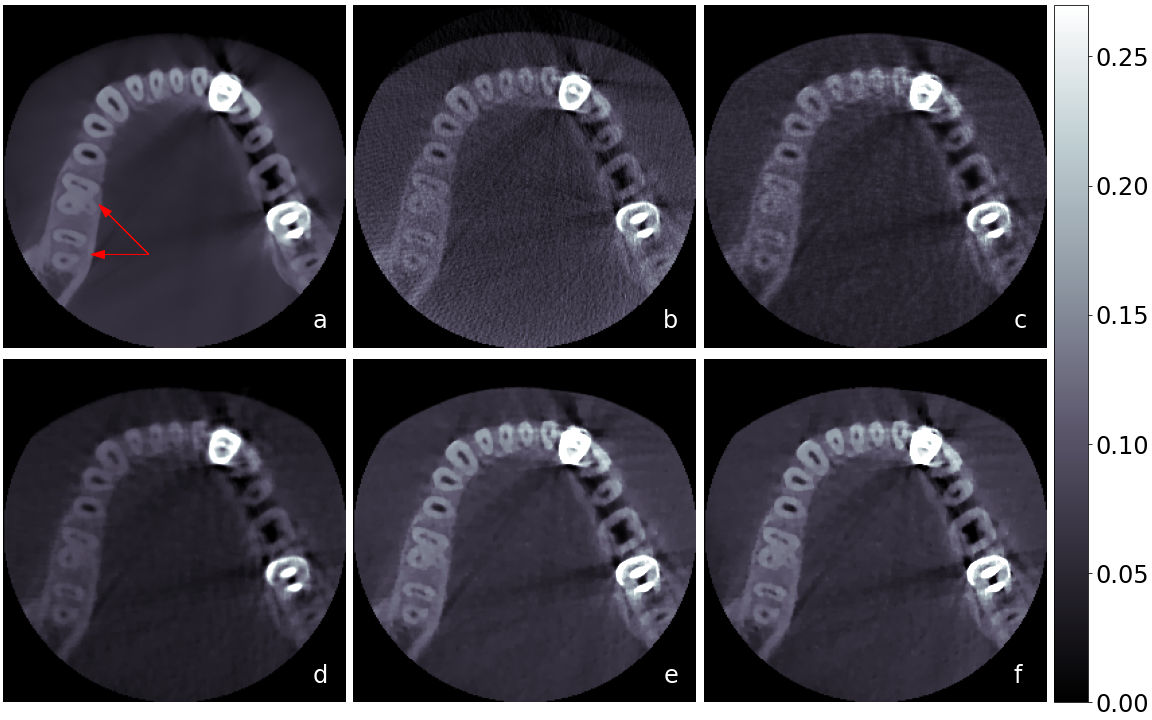}
\caption{Axial slice containing metal from the dental reconstruction, in low-dose. (a) Ground truth, (b) FDK, (c) MLEM, (d) SIRT-TV, (e) MLEM-TV, (f) KL-TV. The red arrows point to the left molars.}
\label{fig:mar_ax120} 
\end{figure}

For the low-dose data, two axial slices are represented in figures \ref{fig:mar_ax180} and \ref{fig:mar_ax120}. Total variation regularization strongly reduces the noise. The background is not as smooth as the normal dose image, but increasing the TV parameter would cause the loss of details that are necessary for a good diagnosis. MLEM-TV and KL-TV images are very similar, with the edge of the teeth sharp, unlike SIRT-TV which appears rather blurred. 

The slice shown in \fref{fig:mar_ax120} contains metal inserts that are known to produce metal artifact in FDK reconstructions. No metal artifact reduction method has been used on any reconstruction. Compared to the previous slice, presence of metal decreases the contrast of the reconstructions, especially in SIRT-TV. We can notice that MLEM, even without regularization, succeeds in reducing the metal artifacts contrary to FDK. In particular, the left molars pointed by a red arrow in \fref{fig:mar_ax120} are more visible. However, the incisors are less sharp on MLEM. 

Figure \ref{fig:mar_sag_zoom} is a zoom on the mandibular left first molar in a coronal slice. Enamel and dentin can be clearly distinguished on all reconstructions. The tooth canals and the bone trabeculae identified respectively by blue and green arrows in \fref{fig:mar_sag_zoom} are more accurately recovered in MLEM-TV and KL-TV images. The mandibular canal (light-blue arrow) is clearly visible on all iterative reconstructions, including MLEM which does not perform denoising.

\begin{figure}[H]
\centering
\includegraphics[width=\textwidth]{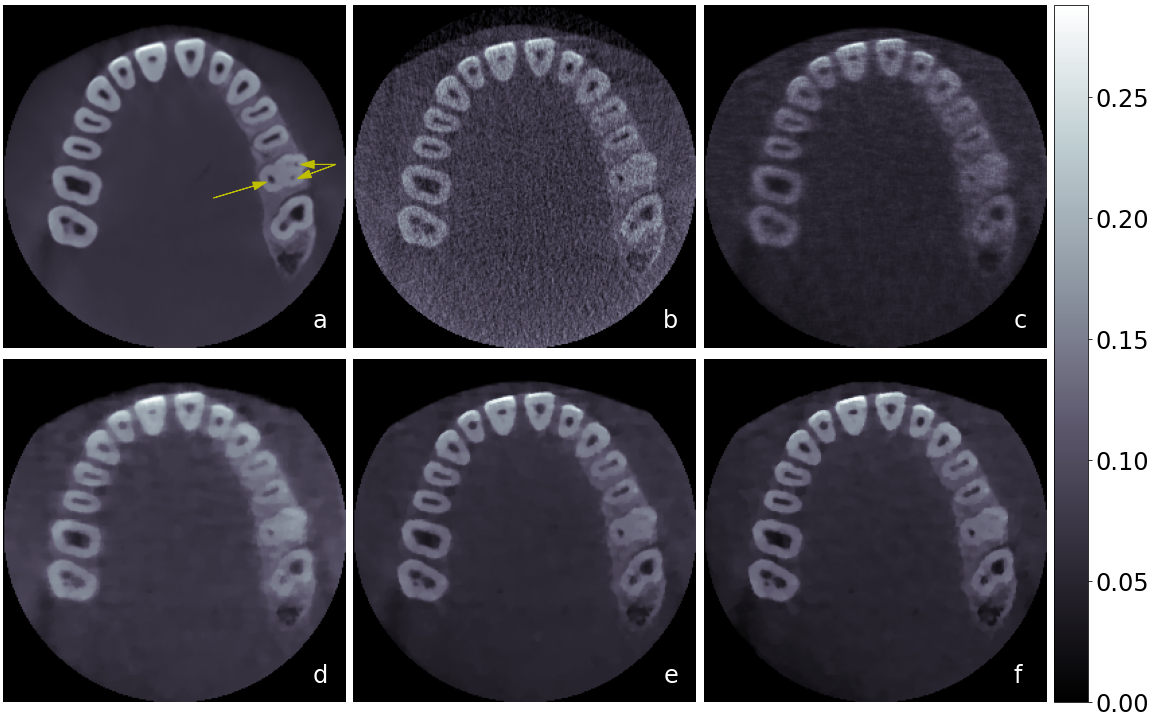}
\caption{Axial slices of the dental reconstructions, in ultra low-dose. (a) Ground truth, (b) FDK, (c) MLEM, (d) SIRT-TV, (e) MLEM-TV, (f) KL-TV. Root canals of the right first molars are identified by the yellow arrows.}
\label{fig:mar2_ax180}
\end{figure}

\begin{figure}[H]\centering
\includegraphics[width=\textwidth]{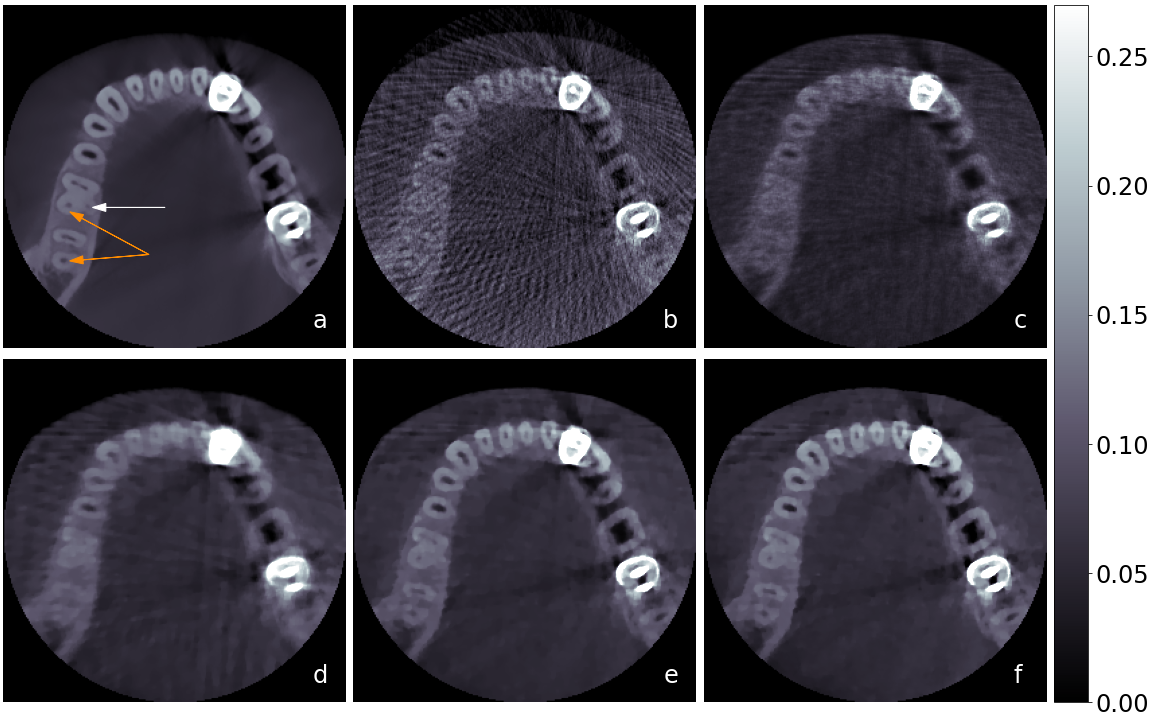}
\caption{Axial slice containing metal of the dental reconstruction, in ultra low-dose. (a) Ground truth, (b) FDK, (c) MLEM, (d) SIRT-TV, (e) MLEM-TV, (f) KL-TV. 
The orange arrows point to root canals erased in MLEM and SIRT-TV reconstruction, and the white one to a canal not reconstructed with any method.}
\label{fig:mar2_ax120} 
\end{figure}

\begin{figure}[H]\centering
\includegraphics[width=0.7\textwidth]{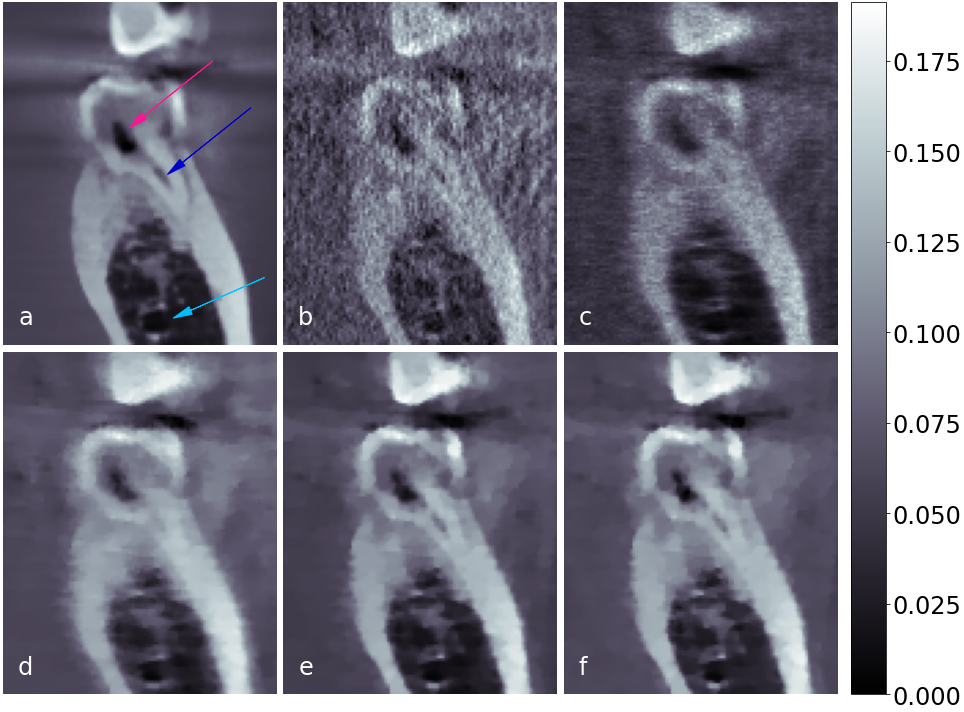}
\caption{Zoom on a tooth in a coronal slice, in ultra low-dose. (a) Ground truth, (b) FDK, (c) MLEM, (d) SIRT-TV, (e) MLEM-TV (f) KL-TV. Pink, blue and light-blue arrows indicate respectively the pulp chamber, the root canal and the mandibular canal.}
\label{fig:mar2_sag_zoom} 
\end{figure}

Figures from \ref{fig:mar2_ax180} to \ref{fig:mar2_sag_zoom} are the reconstructions in ultra low-dose. Overall, the quality of the reconstruction is deteriorated compared to low-dose. In \fref{fig:mar2_ax180}, the noise present in FDK and MLEM reconstructions and the blur of SIRT-TV mask the root canals of the right first molar (yellow arrows). They are made visible by the regularized statistical methods.

The metal artifacts are stronger in the FDK reconstruction compared to the low dose. Comparison with iterative methods based on visual inspection of \fref{fig:mar2_ax120} confirms that iterative methods naturally attenuate these artifacts without any specific correction. The left molars are almost not visible in the FDK reconstruction and are barely distinguishable in MLEM and SIRT-TV images. The root canals pointed by orange arrows were recovered only by MLEM-TV and KL-TV. Some details are also lost in the MLEM-TV and KL-TV reconstruction compared to the low-dose, as the canal at the right of the first left molar (white arrow).

A zoom on the mandibular left first molar already shown in low dose in \fref{fig:mar_sag_zoom} can be seen in \fref{fig:mar2_sag_zoom}. The mandibular canal (light-blue arrow) is visible in the three TV reconstructions. FDK and MLEM gave images where details are lost. The root canal (blue arrow) in the SIRT-TV reconstruction is not visible enough and the pulp chamber (pink arrow) is shrunk. MLEM-TV and KL-TV algorithms reconstruct the most accurate volume, although some details are lost in particular for bone trabeculae.

Reconstructions including regularization clearly outperformed analytical and MLEM reconstructions. Although for the phantom SIRT-TV was comparable with MLEM-TV and KL-TV, in the reconstruction of experimental data the results are blurred and lack contrast. MLEM-TV and KL-TV reconstructions are very similar, with a mutual correlation of over 0.99 in both dose configurations.

The reconstruction times are shown in \tref{tab:marcel_time}. As in the phantom reconstruction, the algorithms using TV regularization are more time consuming than MLEM. The computation times are roughly equivalent for the other three methods in both dose configuration with an advantage for KL-TV. Ranking in terms of speed differs between low-dose and ultra low-dose. This difference comes from the fact that the algorithms do not have the same architecture, so the time complexity with respect to the projections size is different for each of them. 
We conjecture that the computing time is influenced by memory management procedures and might be improved by adequate implementation.

\begin{table}[H]
    \centering
\begin{tabular*}{\textwidth}{l @{\extracolsep{\fill}} lllll}
 & \textbf{FDK} & \textbf{MLEM} & \textbf{SIRT-TV} & \textbf{MLEM-TV} & \textbf{KL-TV} \\
  [-0.45em]  & & \footnotesize(400)  & \footnotesize(400-20)    &  \footnotesize  (400-20)  & \footnotesize (700) \\
\hline
Low-dose & 00:00:06 & 00:30:00 & 01:15:40 & 01:08:40 & 00:58:55 \\
Ultra low-dose & 00:00:03 & 00:27:20 & 00:54:40 & 01:05:44 & 00:55:25 \\
\end{tabular*}
    \caption{Reconstruction times of the experimental data for the different algorithms.}
    \label{tab:marcel_time}
\end{table}

\section{Discussion}

Our results for both simulated and experimental data clearly demonstrate that a significant improvement of image quality can be obtained when regularized models are considered in conjunction with iterative reconstructions. We compared three such algorithms: SIRT-TV, KL-TV and MLEM-TV. While typically classified as an algebraic method, SIRT can also be considered as an EM algorithm maximizing the likelihood for Gaussian data \cite{Yan2011}. Equivalently, this corresponds to the minimization of a quadratic data fidelity term. KL-TV and MLEM-TV both use the Kullback-Leibler distance in the definition of the data fidelity term. Note that minimizing the Kullback-Leibler distance is equivalent to maximize the Poisson likelihood function \cite{Titterington1987}. The Poisson distribution seems better suited to model low-dose data compared to Gaussian distribution, although a more complete model could also account for Gaussian uncertainties at detection. Tests made in 2D with parallel projections which were not shown here suggest that when the Poisson noise is predominant, the algorithms based on the KL distance outperform SIRT-TV while the opposite holds when Gaussian noise is predominant. Our tests have shown better performances for the KL-TV and MLEM-TV algorithms for our experimental data, which supports the need to include Poisson noise when modeling low-dose CBCT.

Although a positive bias is present in zero-valued regions of the MLEM reconstruction, no bias was observed in the experimental data reconstruction. 
Efficient streak artifact reduction was observed with regularized iterative methods. Enhanced by the presence of metal, those artifacts are especially visible in the standard FDK reconstruction. 
Iterative methods reduce them although they remain visible in the low frequencies and cannot be eliminated by an increase in the TV parameter without a significant loss of the details in the images.These artifacts could be further attenuated with a metal artifact reduction algorithm.
 
Table \ref{tab:comparison} summarizes the algorithms used in this study, with their principal strengths and weaknesses.

\begin{table}[H]
\footnotesize \begin{tabular*}{1\textwidth}{l  @{\extracolsep{\fill}} cccccc}
              & Method & Functional                     & Speed & \thead{Artifact \\ reduction}  & \thead{Noise \\ reduction}     & \thead{Detail \\ restoration}      \\
\hline
FDK           & Analytic                & None                                              & ++                     & None      & -         & ++          \\
MLEM          & Likelihood              & $KL(Af,p)$                                               & +                      & +         & -         & -           \\
SIRT-TV       & MAP-EM                  & $\frac{1}{2}\Vert Af - p\Vert^2_2 + \alpha TV(f)$ & -{}-                 & ++        & +         & +           \\
MLEM-TV$\ \,$ & MAP-EM                  & $KL(Af,p) + \alpha TV(f)$                         & -{}-                 & +++       & +         & ++          \\
KL-TV         & PDHG                      & $KL(Af,p) + \alpha TV(f)$                         & -                      & +++       & +         & ++               
\end{tabular*}
\caption{Summary of the five algorithms. We recall the methods on which they are based, the functional to minimize, reconstruction time, artifact reduction capabilities, noise reduction and detail restoration.}
\label{tab:comparison}
\end{table}

In dental CBCT, the distribution of the dose is not uniform through the volume since more dose is given to central region where the region of interest is located. 
The mathematical transcription of the received dose is the sensitivity, defined in equation \eref{eq:sensitivity}. Higher values of the sensitivity are observed at the region of interest and lower values near the boundaries.
Moreover, in dental CBCT, an extension of the volume is necessary because of the truncated projections and leads to even more important discrepancies. This leads to an inhomogeneity in the mathematical projection model not encountered in parallel beam models that has two important consequences on the two regularized statistical methods.
One of them is related to the convergence of the KL-TV and MLEM-TV algorithms. For KL-TV we implemented the preconditioned version of the PDHG algorithm from \cite{Pock2011}. For MLEM-TV we introduced in this work a preconditioned version of the algorithm that prevents numerical divergence when the minimum of the sensitivity is small.
The other consequence concerns the convergence speed and preconditioning allows to improve it for both algorithms. For the original MLEM-TV, the convergence of the TV denoising stage depends on the gradient step $\tau$ in \eref{eq:varphi}, which decreases with the minimum of the sensibility $s_{min}$. With preconditioning we allow this step to vary with the sensitivity, which thus has a faster convergence when denoising the central part of the volume, the one corresponding to the FOV.

To sum up, inhomogeneities in the projection model along with ill-conditioning resulting from projection truncation lead to some challenges, which we were able to overcome by implementing tailored numerical optimization schemes. Not much studies address these issues. We can cite for instance \cite{Stsepankou2012}, where a Poisson data distribution and TV regularization are considered for low-dose CBCT data. The reconstruction is carried out with the MAP-TV algorithm from \cite{Green1990}. However this algorithm is not stable even if numerical convergence can be obtained for small regularization parameters. The upper limit of regularization parameter depends on the minimum of the sensitivity as for MLEM-TV. However in their work, some of the issues we face are circumvented as the experimental data set is composed of complete projections from image-guided radiation therapy.

In terms of computation time, depending on the configuration, MLEM-TV is either comparable or slower than KL-TV. The difference between the two is that KL-TV requires more projection and back-projection operations where MAP-EM performs a full TV denoising at each iteration. These conclusions are different from what was shown in a previous work \cite{Leuliet2021} on 2D reconstructions from parallel projections and on 3D experimental electron microscopy data. This is related once again to the non-uniformity of the sensitivity across the volume. The major part of the computation time in MLEM-TV was spent in the denoising steps. This part was implemented on GPU using the library Cupy from Python. It is likely that some acceleration may be obtained with a more adequate implementation, making TV and reconstruction steps comparable. Using OSEM instead of MLEM could also lead to some acceleration.


The value of the TV parameter influences the denoising quality and the accuracy of the reconstruction. If the parameter is too large, details of the image will be removed, while if it is too small, one will not denoise enough. Hence, the choice of this parameter is decisive for the quality of the reconstruction. A rich literature exists on methods capable to automatically compute the optimal parameter. They are based on the discrepancy principle or on risk estimators. We tried the rule presented in \cite{Ito2011}, where the authors define the regularization parameter by minimizing a functional based on the cost function used in the reconstruction algorithm. If this method had given good results in 2D parallel geometry, in our case, the obtained value was unstable. It has been shown in \cite{Lucka2018} that automatic methods are not necessarily adapted for severely ill-posed problems.
However, we observed in our experiments that with a rather similar configuration (same dose, number of projections, presence of metal or not), the same TV parameter could be kept for all volumes. Thus, it would be possible to determine a fixed number of configurations with their associated parameters.

Iterative methods have some limits. Even if the use of graphics processing units allows to accelerate the iterative 3D reconstructions, they remain rather slow for application in clinical environment and require large memory resources. The TV regularization we applied in this work is largely accepted as a state-of-the-art method. However TV is known to favor piecewise constant areas and produce a "cartoon" effect in the image. Some small structures may be removed and the texture is modified.  

\section{Conclusion}

The main purpose of this work was the evaluation of 3D CBCT reconstruction methods by means of iterative schemes and  TV regularization in dental imaging. The numerical schemes were adapted to the cone-beam geometry and the truncated nature of the projections. We successfully applied four iterative algorithms to 3D phantom and experimental dental data. The test we carried on an ultra low-dose projections set allowed to evaluate the algorithms in a quite extreme situation. The TV regularization clearly improves the quality of the reconstruction. The KL-TV and MLEM-TV algorithms gave the best reconstructions in terms of computation time and reconstruction accuracy.

Compared to standard FDK reconstruction, iterative methods have demonstrated improvements in image quality in reconstruction of low-dose CT data acquired with increased photonic noise due to the lower X-ray tube current.
In this study, we choose to investigate situations where the dose is lowered by reducing the number of acquired projections. With a single acquisition, the normal and the low dose can be obtained simultaneously by subsampling the first one in order to obtain the second. This way, the low dose image can be compared to the reference obtained in normal dose. 
Previous work on other low-dose applications has demonstrated that lowering the current is more advantageous compared to reducing the number of projections \cite{Zhao2014}. Our objective was to investigate the algorithms on experimental data, while still having a reference image of the object. Tests on acquisitions with lower current will be addressed in the future.

Even if the computation times are currently quite large, optimization of the implementation of the algorithms, parallel computing and additional hardware acceleration may lower reconstructions times to the point they become acceptable for clinical practice.
Moreover, these methods have the ability to use the knowledge of the acquisition geometry as well as exact projection and backprojection operations. 

Data-driven methods are an interesting perspective since, once trained, they are expected to provide good quality reconstructions in a few seconds. Nevertheless Deep Learning methods require a large amount of data for training. This is a critical issue in medical imaging, since it may be difficult to obtain a sufficient number of low-dose/high-dose image pairs to train DL networks, especially on patients. Thus, the iterative reconstruction algorithms developed in this work applied to low-dose projections could be used as a surrogate to high-dose images for training DL networks.

\section*{Acknowledgements}
This work was performed within the framework of the LABEX PRIMES (ANR-11-LABX-0063) of Université de Lyon, within the program “Investissements d’Avenir” (ANR-11-IDEX-0007). We acknowledge CARESTREAM Dental for providing us the experimental data.

\section*{Bibliography}

\bibliographystyle{apalike}
\bibliography{JabRef}

\end{document}